\documentclass[usenatbib]{mn2e}   
\pdfoutput=1
\usepackage[pdftex]{graphicx} 
\usepackage{times}
\usepackage{rotate} 
\usepackage{amssymb} 
\usepackage{rotating}
\usepackage{epstopdf}

\title[An IMBH in MRK\,996?]{Searching for an Intermediate Mass
  Black Hole in the Blue Compact Dwarf galaxy MRK\,996} 
\author[Georgakakis et al. ] {A. Georgakakis$^{1}$,
  Y. G. Tsamis$^{2,3}$, B. L. James$^{4}$, A. Aloisi$^4$\\\\
  $^1$National Observatory   of   Athens,   V.   Paulou  \&   I.  Metaxa,   11532,  Greece.\\ 
  $^2$European Southern Observatory, Karl-Schwarzschild-Str. 2, 85748, Garching, Germany.\\ 
  $^3$Department of Physics and Astronomy, The Open University, Walton Hall, Milton Keynes, MK7 6AA, UK.\\ 	
  $^4$Space Telescope Science Institute, 3700 San Martin Drive,  Baltimore, MD 21218, USA.\\
}

\begin{document}
\maketitle  

\begin{abstract}
The  possibility is explored  that accretion  on an  intermediate mass
black hole contributes to the ionisation of the interstellar medium of
the  Compact Blue  Dwarf  galaxy MRK\,996.   Chandra observations  set
tight upper limits (99.7 per  cent confidence level) in both the X-ray
luminosity of  the posited AGN, $L_X(  \rm 2 -  10 \, keV) <  3 \times
10^{40}  \,   erg  \,   s^{-1}$,  and  the   black  hole   mass,  $\la
10^{4}\,\lambda^{-1}  \,  \rm   M_{\odot}$,  where  $\lambda$  is  the
Eddington ratio.  The X-ray  luminosity upper limit is insufficient to
explain  the high ionisation  line [OIV]\,$\rm  25.89\mu m$,  which is
observed in the mid-infrared spectrum  of the MRK\,996 and is proposed
as evidence  for AGN activity.  This indicates  that shocks associated
with  supernovae   explosions  and  winds  of  young   stars  must  be
responsible for  this line.  It is  also found that  the properties of
the  diffuse  X-ray emission  of  MRK\,996  are  consistent with  this
scenario, thereby providing direct  evidence for shocks that heat
the galaxy's interstellar medium and contribute to its ionisation.
\end{abstract}


\section{Introduction}\label{sec_intro}

Understanding  how supermassive black  holes (SBH)  grow and  how they
relate  to the  formation of  their host  galaxies are  challenges for
modern   astrophysics.    Black  holes   with   masses  $\rm   \approx
10^{9}\,M_{\odot}$ have now been observed out to $z\approx6$, when the
Universe   was   only   about  1\,Gyr   old   \citep[e.g.][]{Fan2006}.
Simulations can  produce such monsters at early  epochs by postulating
violent high  accretion rate events  triggered by mergers of  gas rich
galaxies    \citep[e.g.][]{Volonteri_Rees2005}.    These   simulations
however, have to assume seed black holes, which form very early in the
Universe ($z\approx20$)  and which will  eventually grow to  masses as
high as $\rm 10^{9}\,M_{\odot}$  via accretion.  The mass distribution
of the seed black holes  is an important parameter of the simulations.
The  remnants  of  the  first   stars  that  formed  in  the  Universe
(Population III) are believed to produce black holes with masses up to
$10^3\,M_{\odot}$
\citep{Abel_Bryan_Norman2000,Bromm_Coppi_Larson2002}.   In  this  case
however, accretion well in excess of the Eddington rate is required to
grow  those seeds  to  $\rm \approx  10^{9}\,M_{\odot}$  in less  than
1\,Gyr     \citep[e.g.][]{Volonteri_Rees2005,    Begelman2006}.     An
alternative  scenario for  the formation  of seed  black holes  is the
direct collapse of pre-galactic gas discs \citep{Lodato_Natarajan2006,
  Lodato_Natarajan2007}.  This model  produces black holes with masses
up  to  $10^{6}\,M_{\odot}$, orders  of  magnitude  more massive  than
Population III stars,  thereby significantly relaxing the requirements
for  super-Eddington accretion.  A  powerful probe  of the  seed black
hole  population at  early epochs  are intermediate  mass  black holes
($\rm  \la  10^{6}\,M_{\odot}$)  in  the  local  Universe,  which  are
expected to  reside in low mass  galaxies \citep[i.e.  $M_{BH}-\sigma$
  relation;][]{Ferrarese2000,        Gebhardt2000,        Greene2010}.
\cite{Volonteri2008} for example, argue that the black hole occupation
number of  dwarf galaxies  can constrain models  of the origin  of the
seed black holes in the  early Universe, e.g.  Population III stars vs
direct collapse of pre-galactic gas discs.

The  dynamical signatures  of moderate  size black  holes are  hard to
detect with  current instrumentation.  As  a result searches  for such
objects in nearby  galaxies have turned to the  easier but less direct
approach of  finding AGN  in small or  bulgeless galaxies,  i.e. black
holes during their  active stage.  NGC\,4395 and POX\,52  are the best
studied examples  of nearby broad-line AGN with  estimated black holes
mass              of             about             $10^{5}\,M_{\odot}$
\citep{Filippenko2003,Barth2004,Peterson2005}.   Furthermore,  optical
spectroscopic  studies have  recently  found evidence  for type-2  AGN
activity  associated with intermediate  mass black  holes in  the late
type   spirals  NGC\,1042  \citep{Seth2008,   Shields2008},  UGC\,6192
\citep{Barth2008} and NGC\,3621 \citep{Barth2009}.  Systematic efforts
to compile samples of moderate size black holes have also been carried
out by \cite{Greene_Ho2004, Greene_Ho2007} and \cite{Dong2007}.  Those
studies used the SDSS spectroscopic database to identify galaxies with
AGN  signatures in  their  optical spectra  and  estimated black  hole
masses of  $<10^{6}\,M_{\odot}$. The hosts of those  AGN are typically
late-type galaxies with stellar masses $<10^{10}\,M_{\odot}$.

A  special  class of  low  mass  galaxies  which have  attracted  much
attention  over  the years  are  Blue  Compact  Dwarfs (BCDs).   These
systems are exceptional because  of their low metallicities (1/40--1/3
solar) and  high specific star-formation rates, which  indicate one of
the first major episodes of star-formation.  BCDs are therefore unique
laboratories for  testing theories of galaxy  formation. They resemble
the first galaxies  that formed in the early Universe  out of gas with
near-primordial  metallicity,  and  which  according to  the  standard
hierarchical  formation paradigm  are the  building blocks  of massive
galaxies. An  important recent  development in the  study of  BCDs has
been the suggestion that  they harbor AGN associated with intermediate
mass black holes.  If true,  this would offer the opportunity to study
in detail the parallel formation of the first stars and the seed black
holes in systems that resemble proto-galaxies.

Evidence  for AGN  activity in  BCDs inlcude,  the incidence  of dense
ionised  cores ($\rm  >10^4 \,cm^{-3}$),  broad Balmer  emission lines
with     widths     in     excess    of     $\rm     2000\,km\,s^{-1}$
\citep{Izotov2007,Izotov_Thuan2008},  high  ionisation narrow  optical
emission lines  suggesting the presence  of a hard  ionising continuum
\citep{Thuan_Izotov2005,James2009}.   The  evidence  for AGN  in  BCDs
however, is far from conclusive.   Broad and high excitation lines can
be  due to  stellar  activity, e.g.   shocks  in fast  winds of  young
massive      stars     \citep[e.g.][]{Izotov2007,Thuan2008,James2009}.
Unfortunately,    optical    emission    line   diagnostic    diagrams
\citep[e.g.][]{Baldwin_Phillips_Terlevich1981, Kewley2001}, which have
traditionally been used  to separate stellar photoionization-dominated
from  AGN-excited sources, break  down at  the low  metallicity regime
\citep{Groves2006},  and hence cannot  be applied  to BCDs.   The most
reliable method for  confirming or refuting the presence  of an AGN in
any galaxy, irrespective of  metallicity, are X-ray observations.  The
detection of an X-ray point source  at the nuclear regions of a galaxy
is strong evidence for accretion on a massive compact object.

In this paper we explore the  X-ray properties of MRK\,996, a BCD that
is among the prime candidates  for hosting an AGN. Mid-IR spectroscopy
of MRK\,996 with Spitzer IRS shows the [OIV]\,25.89$\, \rm \mu m$ line
with an ionisation potential of 54.9\,eV, indicating the presence of a
hard  continuum  \citep{Thuan2008},  which  could be  produced  by  an
AGN. Optical integral field spectroscopy by VLT/VIMOS have revealed an
extremely dense nuclear region  (electron density $\rm \approx 10^7 \,
cm^{-3}$), much denser than  in typical BCDs \citep{James2009}.  Also,
a very  red point source ($J-K=1.8$)  was found at the  nucleus of the
galaxy,  while  photometry with  Spitzer  indicates  hot  dust in  the
central galaxy regions and a mid-IR Spectral Energy Distribution (SED)
similar   to  that   of  Seyferts   \citep{Thuan2008}.    New  Chandra
observations of MRK\,996  are presented in this paper  and are used to
search  for  AGN signatures  in  this  galaxy.   We adopt  a  distance
of 22.3\,Mpc for MRK\,996, which corresponds to a velocity of $\rm
1642\,km\,s^{-1}$ (James et  al. 2009) and $\rm H_{0} =  73.5 \, km \,
s^{-1} \, Mpc^{-1}$.

\section{Data processing}

MRK\,996 was observed for a  total of 50\,ks with the ACIS-S (Advanced
CCD  Imaging Spectrometer) camera  aboard Chandra  in August  3, 2009.
The ACIS-S was  operated in very faint mode and  the target was placed
at the nominal aimpoint of the S3 CCD. The data were reduced using the
Chandra  Interactive Analysis of  Observations (CIAO)  package version
4.1.2 and  the methodology  described in \cite{Laird2009}.   The first
step was  to produce  the level 2  event file  from the raw  data. Hot
pixels and  cosmic ray  afterglows were flagged.   The charge-transfer
inefficiency  (CTI)  and  time-dependent  gain corrections  were  then
applied and the ACIS pixel randomization was removed.  As MRK\,996 was
observed  in very faint  mode, the  ACIS particle  background cleaning
algorithm was also applied. The  data were inspected for high particle
background  periods and  it was  found  that they  were not  affected.
Images  were constructed  in  four energy  bands 0.5-7.0\,keV  (full),
0.5-2.0\,keV    (soft),   2.0-7.0\,keV    (hard)    and   4.0-7.0\,keV
(ultra-hard).  Sources  were detected using  the methodology described
in \cite{Laird2009}.   The final source list consists  of sources with
Poisson probabilities $<4\times10^{-6}$ in  at least one of the energy
bands defined above.  Source free background maps and sensitivity maps
were also constructed as described in \cite{Georgakakis2008_sense}.

The  astrometry of  the  Chandra  data was  refined  by comparing  the
positions of X-ray  sources with those of optical  galaxies in the HST
Wide  Field  Planetary  Camera  2 (WFPC2)  observations  presented  by
\cite{Thuan1996}.   The  astrometric  accuracy  of the  HST  image  is
0.25\,arcsec.  Excluding sources within the optical radius of MRK\,996
(0.3\,arcmin) obtained  from the  Third Reference Catalogue  of Bright
Galaxies \citep[RC3;][]{rc3},  there are  only 2 X-ray  sources within
the HST/WFPC2 field of view.  Both of them are associated with optical
galaxies. Comparison  of the optical and X-ray  source positions shows
small  systematic offsets  of  $\rm \delta  RA=0.25$\,arcsec and  $\rm
\delta  DEC=0.05$\,arcsec,  which were  applied  to  the X-ray  source
catalogue and images.

\begin{figure*}
\begin{center}
\includegraphics[height=0.7\columnwidth]{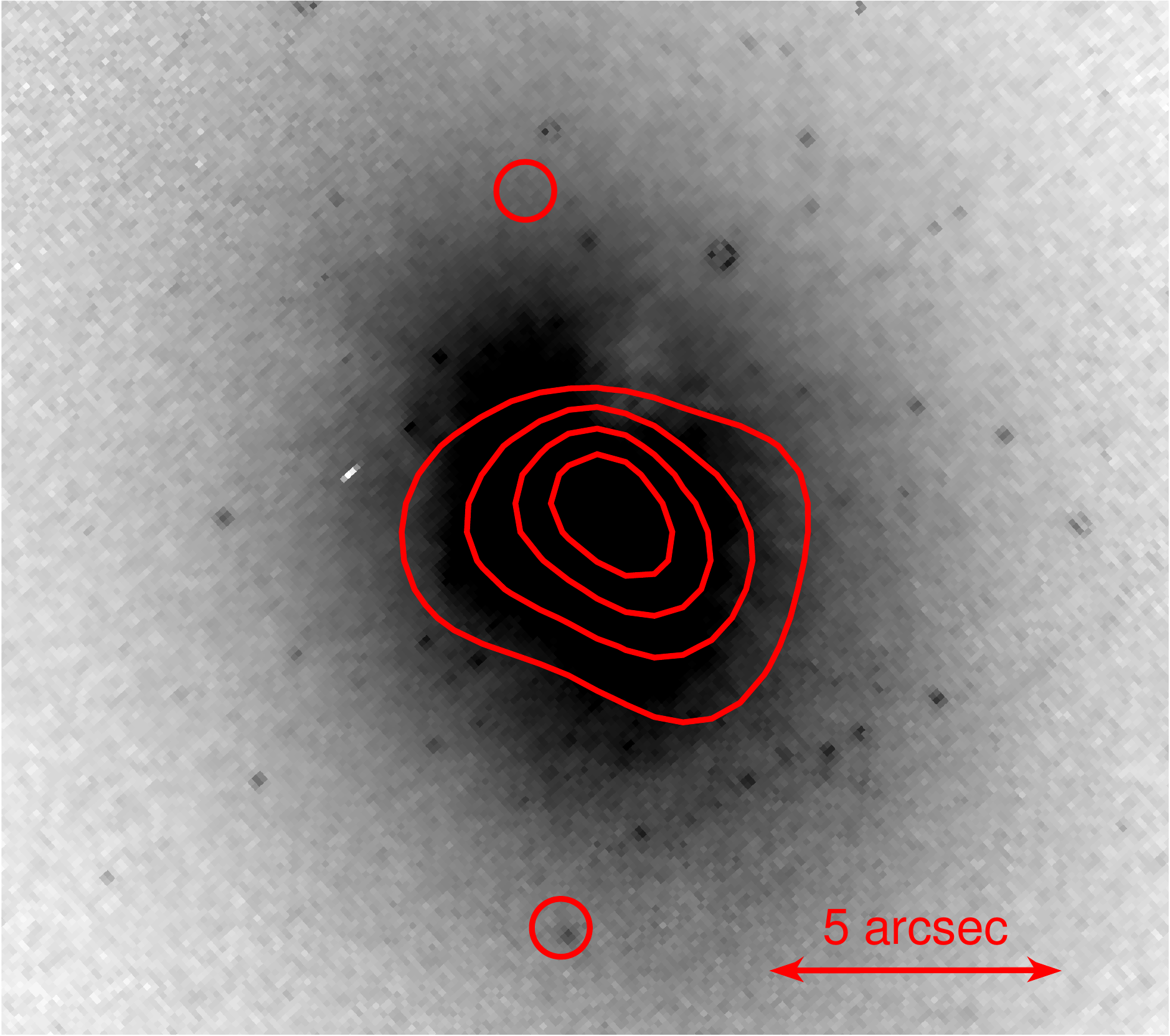}
\includegraphics[height=0.7\columnwidth]{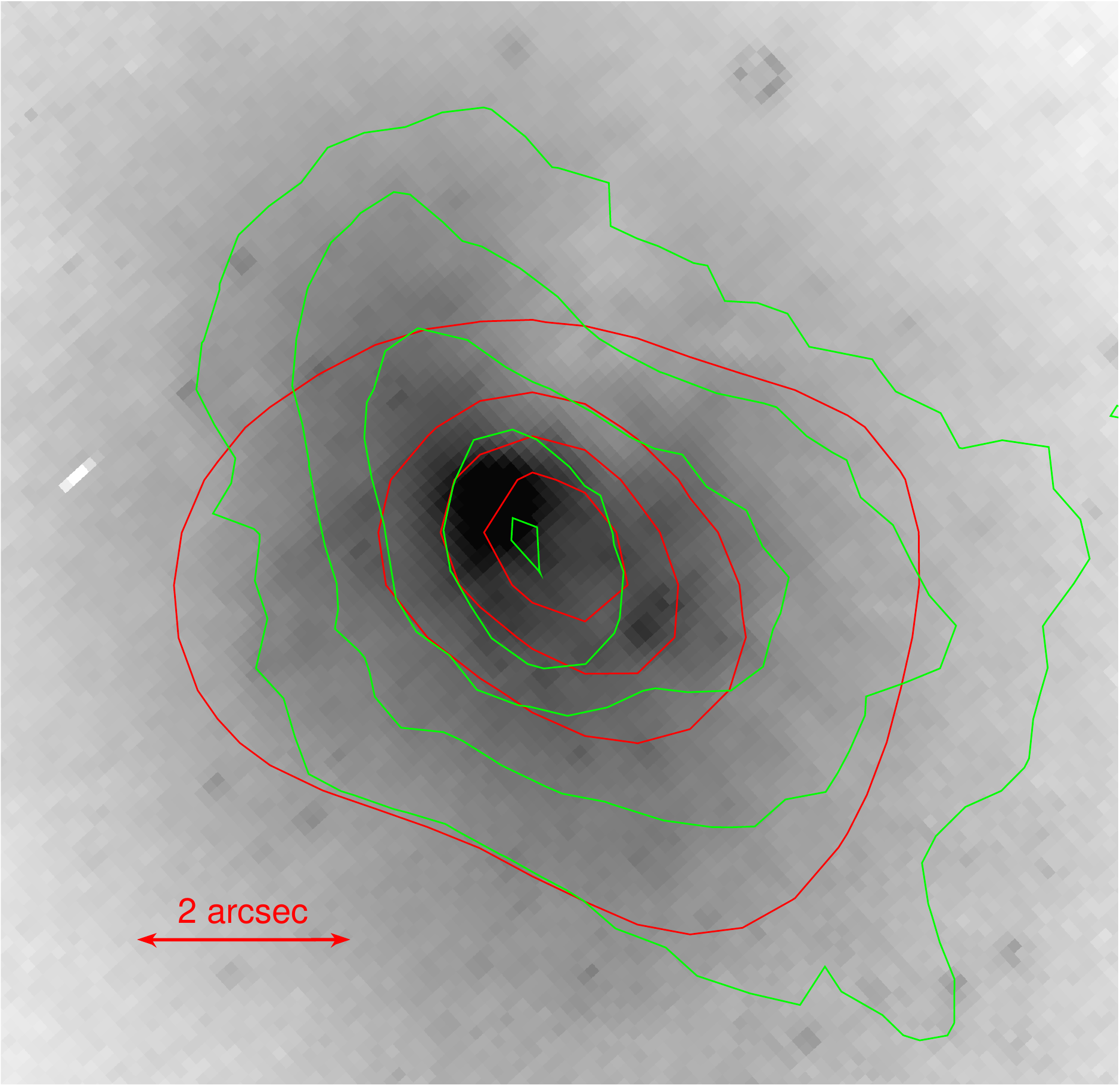}
\end{center}
\caption{{\bf Left:} HST F569W  filter image of MRK\,996.  The circles
  are 0.5\,arcsec in radius and  mark the positions of the X-ray point
  sources  detected   in  the  Chandra   observations.   The  contours
  correspond to the diffuse X-ray  emission of MRK\,996.  North is up,
  East is left.  {\bf Right:}  Zoom in the nuclear regions of MRK\,996
  to compare the  spatial distributions of the hot  X-ray emitting gas
  versus the optically  detected H\,II gas.  The greyscale  is the HST
  F569W filter image, the red contours correspond to the diffuse X-ray
  component  (same  as contours  on  the  left  image) and  the  green
  contours  are the narrow  component of  the H$\alpha$  emission line
  from  the  integral field  spectroscopic  observations described  by
  James et al.  (2009).  }\label{fig_mrk996_ps}
\end{figure*}

\section{X-ray point sources}

At the  Poisson probability  threshold of $4\times10^{-6}$,  two X-ray
point  sources are  detected  within the  optical  radius of  MRK\,996
(0.3\,arcmin).   Figure 1 (left)  overplots the  positions of  the two
sources on the HST image of MRK\,996.  None of them is associated with
the nucleus of the galaxy. Relative to the optical centre of MRK\,996,
one of the point sources  lies $\approx7$\,arcsec to the south and the
other $\approx6$\,arcsec  to the north.  The  southern X-ray detection
is likely associated with one of the globular clusters of MRK\,996. It
lies  $\approx0.15$\,arcsec from an  optical point  source on  the HST
image  of the  galaxy identified  as a  globular cluster  by  Thuan et
al. (1996).

The  X-ray  fluxes and  luminosities  of  the  two point  sources  are
estimated  from their X-ray  spectra.  The  {\sc specextract}  task of
CIAO  was employed  to extract  the counts  at the  positions  of each
source,  estimate  background  spectra,  generate  the  Redistribution
Matrix Function  (RMF) and the  Ancillary Response File  (ARF). Source
spectra were grouped to have at least one count per bin.  The spectral
analysis  was carried out  within the  XSPEC v12  package. As  the net
counts are  low for both  sources (see Table \ref{tab_xspec}),  we fit
the data  with a simple  power-law model absorbed  by cold gas  in our
Galaxy.  A  Galactic hydrogen  column density of  $\rm N_H =  4 \times
10^{20}   \,   cm^{-2}$   was    adopted   using   the   HI   map   of
\cite{Kalberla2005}.    In    XSPEC   terminology,   our    model   is
$\mathtt{wabs} \times \mathtt{po}$,  where $\mathtt{wabs}$ is the cold
absorption  from our Galaxy  and $\mathtt{po}$  is the  power-law. The
derived   parameters  are  listed   in  Table   \ref{tab_xspec}.   The
luminosities  of the  two sources  are  typical of  X-ray binaries  in
galaxies.

We assess the  probability that the two off-nuclear  X-ray sources are
background AGN.   We use the  0.5-10\,keV $\log N-\log S$  relation of
\cite{Georgakakis2008_sense}  to  estimate   the  expected  number  of
sources  brighter than  $\rm 10^{-15}  \,  erg \,  s^{-1} \,  cm^{-2}$
within  a radius  of  $\approx7$\,arcsec from  the  optical centre  of
MRK\,996.   The  flux  limit  above  corresponds  to  the  approximate
0.5-10\,keV flux of  the faintest of the two  sources.  The radius cut
is  the maximum angular  distance of  the two  point sources  from the
galaxy  centre.  The expected  number of  background sources  is 0.12.
The    Poisson   probability    of   two    background    AGN   within
$\approx7$\,arcsec    of    the    MRK\,996   nuclear    regions    is
$1.8\times10^{-4}$.   We conclude  that both  X-ray point  sources are
likely to be associated with MRK\,996.

\section{Diffuse emission}\label{sec_diffuse}

The nuclear  regions of  the MRK\,996 are  dominated by  diffuse X-ray
emission. We explore the morphology of the X-ray gas by first removing
point sources and then adaptively  smoothing the full band X-ray image
using  the {\sc  cmooth}  task of  CIAO.   The X-ray  contours of  the
smoothed  image are  overplotted  on the  HST  observations in  Figure
\ref{fig_mrk996_ps}.  The  peak of the diffuse X-ray  emission lies in
between the nuclear  star-forming region and a point  source in the SW
which was  proposed to  be a young  star cluster  by \cite{Thuan1996}.
There is also tentative  evidence, limited by small photon statistics,
that the distribution of the diffuse component is more extended in the
southern direction. This is interesting,  if it is confirmed by deeper
X-ray observations,  as the spatial distribution  of globular clusters
of MRK\,996 also appears to  be assymetric in the same direction. Most
of them are found in the south of the galaxy \citep{Thuan1996}. Figure
\ref{fig_mrk996_ps} (right)  also shows that  the spatial distribution
of the diffuse X-ray component and the H\,II ionised gas are similar.

\begin{figure}
\begin{center}
\includegraphics[height=0.9\columnwidth,
  angle=270]{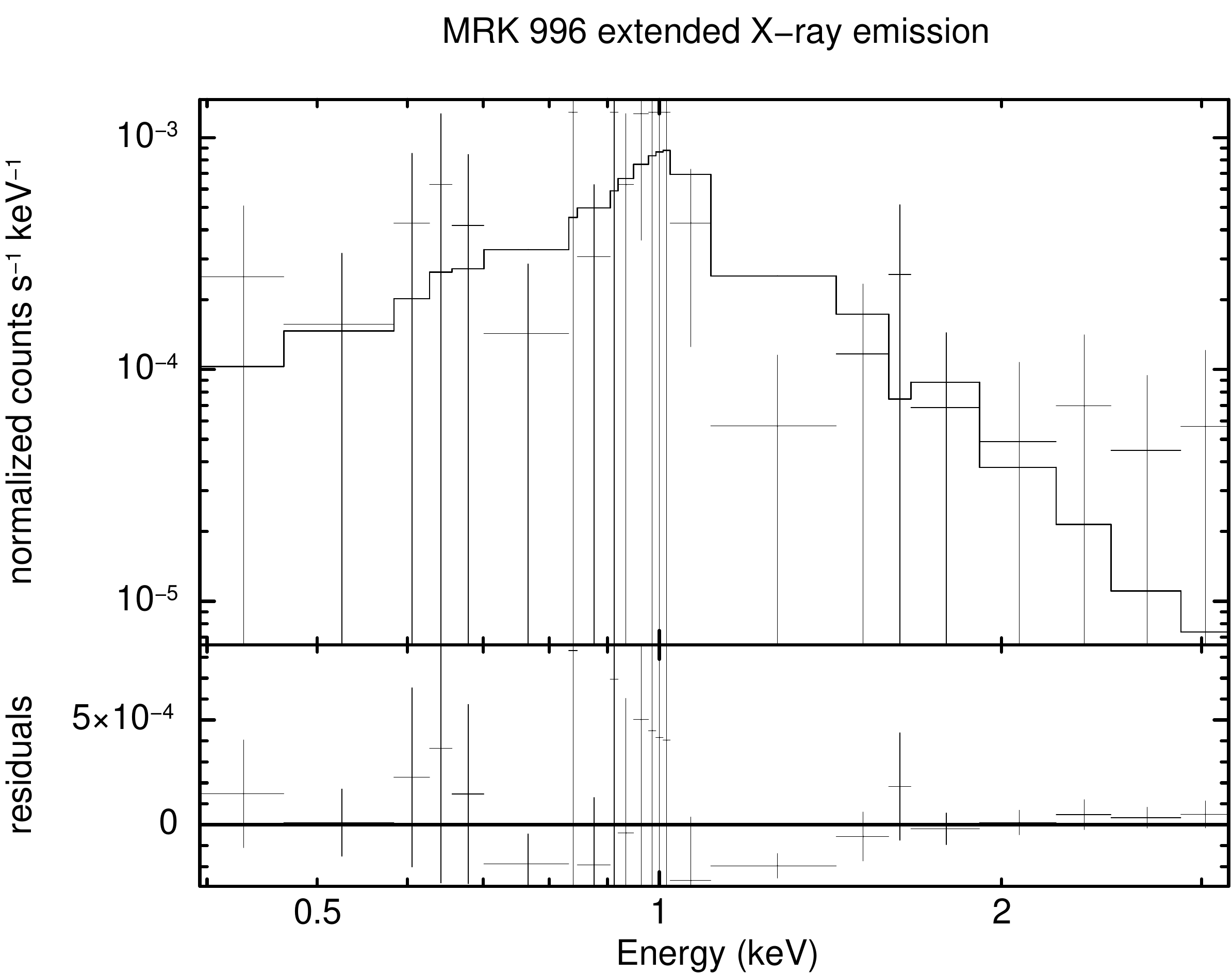}
\end{center}
\caption{Spectrum of the diffuse  X-ray emission of MRK\,996. The best
  fit thermal plasma model is shown with the histogram.  The residuals
  of the model are also plotted.  }\label{fig_xgas}
\end{figure}

The X-ray spectrum  of the diffuse component has  been extracted using
{\sc specextract} task of CIAO which  is the proposed tool in the case
of extended  sources.  The extraction  aperture has a radius  of about
2\,arcsec and  encompasses most of the diffuse  component photons. The
background  spectrum was determined  from nearby  source-free regions.
The   spectrum   was   fit   with   a   hot   thermal   plasma   model
\cite{Raymonmd_Smith1977} which is absorbed  by cold gas in our Galaxy
with an HI column density of  $\rm N_H = 4 \times 10^{20} \, cm^{-2}$.
The  X-ray  spectrum  and  the  best-fit model  are  shown  in  Figure
\ref{fig_xgas}.  Table \ref{tab_diffuse}  lists the hot gas parameters
estimated from the best-fit model  to the data. The temperature of the
thermal  plasma, about 1\,keV  ($\approx \rm  10^{7}\, K$),  is higher
than  the  typical  temperatures  of  the diffuse  hot  gas  in  dwarf
starbursts \citep[e.g.][]{Ott2005} and other BCDs \citep{Thuan2004}.

\begin{table*}
\begin{center}

\caption{MRK\,996  point source spectral  parameters. Columns  are (1)
  CXO ID, (2) total counts  within the extraction region, (3) best-fit
  photon index  of the power-law  model, (4) c-statistic value  of the
  fit and  degrees of freedom, (5)  flux in the  0.3-10\,keV band, (6)
  X-ray luminosity in the 0.3-10\,keV band.  } \label{tab_xspec}

\begin{tabular}{l cc ccc}\hline
ID & Total counts & $\Gamma$  & c-stat/dof & $f_X(\rm 0.3 - 10 \,keV)$
& $L_X(\rm 0.3 - 10 \,keV)$ \\  & & & & ($\rm 10^{-14} \, erg \,s^{-1}
\, cm^{-2}$)& ($\rm erg\,s^{-1}$) \\

\hline  CXO\,J012735.6-061930   &  74   &  $4.0\pm0.6$  &   33.7/24  &
$2.0^{+1.0}_{-0.6}$     &    $1.2^{+0.6}_{-0.4}     \times    10^{39}$
\\  CXO\,J012735.5-061943  &  11   &  $2.5^{+2.1}_{-1.6}$  &  8.9/9  &
$0.3^{+0.2}_{-0.1}$ & $1.8^{+1.2}_{-0.6} \times 10^{38}$ \\ \hline
\end{tabular}
\end{center}
\end{table*}

\begin{table}
\begin{center}

\caption{ The gas parameters for  the thermal plasma model used to fit
  the spectrum  of the diffuse  emission of MRK\,996.   The parameters
  are estimated for  a metallicity $Z_{opt}=0.5\,Z_{\odot}$ determined
  by James  et al.   2009 from the  optical emission line  spectrum of
  MRK\,996  and  a volume  filling  factor  $f_{v}=1$.  The  parameter
  values  can  be  converted   to  lower  filling  factors  and  other
  metallicities using  the scaling relation given in  the Table, where
  $\xi=Z/Z_{opt}$.   The  listed   parameters  are:  (1)  best  plasma
  temperature  in keV, (2)  the normalisation  of the  best-fit model,
  which equals $10^{-14}\, (4 \, \pi \, D^{2})^{-1} \, \int n_e \, n_H
  \, dV$,  where $D$ is the  distance to MRK\,996 in  cm, $n_e$, $n_H$
  are  the electron  and proton  densities in  $\rm cm^{-3}$,  (3) the
  plasma  temperature in K,  (4) The  0.3-10\,keV X-ray  luminosity in
  erg/s, (5) the mean electron density in $cm^{-3}$ estimated from the
  normalisation of the best-fit model assuming the proton and electron
  densities are equal, (6) pressure  of the hot gas estimated from the
  relation    $P/k=2\,n_e\,T$,   (7)    the   gas    thermal   energy,
  $E_{th}=3\,n_e\,V  \,kT$,  where $V$  is  the  volume  of the  X-ray
  emitting   region,   which   is   assumed   to  be   a   sphere   of
  2\,arcsec=214\,pc radius  at the  source distance (i.e.   similar to
  the  extraction  region  of   the  X-ray  spectrum  of  the  diffuse
  component),  (8)  total mass  of  the  hot  gas estimated  from  the
  relation $M_{hot}=n_e  \, m_{p} \,  V$, where $m_{p}$ is  the proton
  mass, (9)  gas cooling time, $t_{cool}=E_{th}/L_{X}$  and (10) X-ray
  emissivity estimated from the relation $\Lambda_X=L_X/(norm \times 4
  \, \pi \, D^{2} \times 10^{14})$.  }\label{tab_diffuse}

\begin{tabular}{lc}\hline
Parameter  &  Value\\   \hline  $kT$  (keV)  &  $1.1\pm^{+0.4}_{-0.2}$
\\\\ $norm$ ($10^{-6}$) & $1.8^{+0.9}_{-0.7}$ \\\\

$T$ ($\rm 10^{7}\,K$) & $1.3\pm^{+0.5}_{-0.2}$ \\\\

$L_{X}(\rm          0.3-10\,keV)$         ($10^{38}$\,erg/s)         &
$1.1\pm^{+0.6}_{-0.4}$\\\\

$<n_e>$     [$\times    (f\,\xi)^{-0.5}$]     ($\rm     cm^{-3}$)    &
$0.10^{+0.03}_{-0.02}$ \\\\

$P/k$  [$\times  (f_{v}\,\xi)^{-0.5}$]  ($\rm  10^{6}  K\,cm^{-3}$)  &
$2.7^{+1.4}_{-0.7}$\\\\

$E_{th}$ [$\times f_{v}^{0.5}\, \xi^{-0.5}$]  ($\rm 10^{53} \, erg$) &
$5.2^{+3.6}_{-1.9}$\\\\

$M_{hot}$   [$\times  f_{v}^{0.5}\,   \xi^{-0.5}$]  ($\rm   10^{5}  \,
M_{\odot}$) & $0.8^{+0.3}_{-0.2}$\\\\

$t_{cool}$ [$\times  f_{v}^{0.5}\, \xi^{-0.5}$]  ($\rm 10^8 \,  yr$) &
$1.5^{+1.1}_{-0.4}$\\\\

$\Lambda$    ($\rm   10^{-23}   \,erg    \,s^{-1}   \,    cm^{3}$)   &
$1.0^{+0.7}_{-0.5}$\\ \hline
\end{tabular}
\end{center}
\end{table}

\section{Discussion}

\subsection{Is there an AGN in MRK\,996?}

One of the motivations for the Chandra observations of MRK\,996 was to
explore the possibility that an  AGN is responsible for the excitation
of   the  mid-IR  [OIV]\,25.89$\,   \rm  \mu   m$  line   reported  by
\cite{Thuan2008}.   Below  about 2\,keV  the  nuclear  regions of  the
galaxy are dominated by diffuse  hot gas, which renders the search for
a point  source in the soft  and full bands difficult.   The hard band
(2-7\,keV)  is less contaminated  by the  diffuse X-ray  emission.  We
therefore  estimate   a  conservative  upper  limit   (99.7  per  cent
confidence level) for the flux of a point source at the optical centre
of MRK\,996  assuming that  the observed counts  at that  position are
from a nuclear X-ray point source.

For the estimation of the upper limits we adopt a Bayesian methodology
similar  to that  described in  \cite{Laird2009}.  For  a  source with
observed number of total counts $N$ (source and background) within the
70  per  cent EEF  (Encircled  Energy  Fraction)  radius and  a  local
background value $B$ the probability of flux $f_X$ is

\begin{equation}\label{eq_source} P(f_X, N) = \frac{T^N \, e^{-T}}{N!},
\end{equation}

\noindent where  $N=S+B$ and $S =  f_X \times t_{exp}  \times C \times
\eta$.   In the  last equation  $t_{exp}$ is  the exposure  time  at a
particular   position  after   accounting  for   instrumental  effects
(i.e. exposure map),  $C$ is the conversion factor  from flux to count
rate which depends on the adopted model for the spectrum of the source
and $\eta$ is the encircled energy fraction, i.e. 0.7 in our case. The
integral of  $P(f_X, N)$  gives the probability  that the flux  of the
source is  lower than $f_{X,U}$,  i.e. the confidence interval  CL for
the flux upper limit

\begin{equation}\label{eq_limit} \int_{f_{X,L}}^{f_{X,U}} P(f_X, N) \,
df_x=CL.
\end{equation}

\noindent The lower limit of the integration is $f_{X,L}= \rm 10^{-18}
\, erg \, s^{-1} \,  cm^{-2}$.  Equation \ref{eq_limit} is then solved
numerically to estimate  the flux upper limit that  corresponds to the
confidence  interval of 99.7  per cent.   For the  flux to  count rate
conversion factor we adopt  two extreme values.  The first corresponds
to  a  power-law with  $\Gamma=1.9$  and  represents  the case  of  an
unabsorbed AGN.   The second is  for a reflection  dominated spectrum,
which   is  approximated   by  the   {\sc  pexrav}   model   of  XSPEC
\citep{Magdziarz1995} and  represents the case of a  Compton Thick AGN
($\rm N_H \ga 10^{24} \, cm^{-2}$).   The 99.7 per cent upper limit is
$f_X(\rm  2-10 \,  keV)  = 6\times10^{-15}  \rm  \, erg  \, s^{-1}  \,
cm^{-2}$ for the $\Gamma=1.9$ model and $10^{-14} \rm \, erg \, s^{-1}
\,  cm^{-2}$ for  the reflection  dominated spectrum.   The 2--10\,keV
luminosity   99.7   per    cent   upper   limits   are   respectively,
$3.5\times10^{38}$ and $6 \times10^{38}\rm  \, erg \, s^{-1}$.  In the
case of a reflection dominated  spectrum, we assume that the intrinsic
luminosity of  the source is  50 times higher ($3\times10^{40}  \rm \,
erg \, s^{-1}$), as the  reflected emission is believed to represent 2
per  cent  of  the  direct  component  \cite[e.g.][]{Gilli2007}.   The
corresponding  upper   limits  for  the  black  hole   mass  are  $\la
100\,\lambda^{-1} \rm  \,M_{\odot}$ in the case of  no obscuration and
$\la 10^4 \, \lambda^{-1} \rm \,M_{\odot}$ for Compton thick obscuring
clouds, where $\lambda$ is the Eddington ratio.  In this calculation a
bolometric  to   hard  X-ray  luminosity   ratio  of  35   is  adopted
\citep{Elvis1994}.

The  luminosity upper  limits above  should be  compared with  the AGN
luminosity required to explain the  [OIV]\,25.89$\, \rm \mu m$ line in
the mid-IR  spectrum of MRK\,996 reported  by \cite{Thuan2008}.  Those
authors modeled this line with  an AGN ionization spectrum of the form
$f_\nu  \propto  \nu^{-1}$ and  estimated  a  total  rate of  ionizing
photons $\log  Q(H)=51.125$.  Assuming the photons  are distributed in
the energy interval 1--1000\,Ry,  this rate translates to a 2--10\,keV
intrinsic AGN  luminosity of $4.5\times10^{40} \, \rm  erg \, s^{-1}$.
This is 2\,dex higher than the 99.7 luminosity upper limit in the case
of an  unabsorbed AGN.   This indicates that  an unabsorbed  or mildly
absorbed active  SBH cannot be  responsible for the ionisation  of the
[OIV]\,25.89$\, \rm \mu m$ line.  If the central engine in MRK\,996 is
absorbed by Compton Thick material and the X-ray emission is dominated
by the  reflected component, the  estimated 99.7 per cent  upper limit
for the  intrinsic luminosity  of the AGN  is still lower  compared to
that  required to  explain the  [OIV]\,25.89$\, \rm  \mu m$  line.  We
conclude that  the X-ray data provide evidence  against AGN ionisation
for the  [OIV]\,25.89$\, \rm  \mu m$ line  at a significance  level of
about 99.7 per cent.

\subsection{Origin of the diffuse component}

An alternative explanation for the  [OIV]\,25.89$\, \rm \mu m$ line is
ionisation  via  shocks   \citep{Thuan2008}  associated  with  stellar
outflows and supernovae explosions.  Shock excitation has already been
proposed to partly explain the broad component of the optical emission
lines  in   MRK\,996  \citep{James2009}.   The   shocks  generated  by
starburst  winds  are  also  known  to  carve  bubbles  of  hot  ($\rm
10^{6}-10^{7}\,K$) and overpressured  gas in the inter-stellar medium,
which can be observed at soft X-rays \citep[e.g.][]{Ott2005}.

We explore the  possibility of a superbubble for  the observed diffuse
X-ray emission by  comparing the observed X-ray luminosity  of the hot
gas,  as  determined  from  the  X-ray spectral  analysis  of  section
\ref{sec_diffuse},  with  the  predictions  of the  bubble  models  of
\cite{Castor_McCray_Weaver1975}, \cite{Weaver1977}, \cite{MacLow1988}

\begin{equation}
L_X= \int_{0}^{R_{max}} n(r)^2 \, \Lambda_X(T,Z) \,dV,
\end{equation}

\noindent where $\Lambda_X(T,Z)$ is the X-ray volume emissivity of the
gas  and  $n(r)$  is  the   radial  density  profile  of  the  bubble.
Substituting   in  the   integral   above  the   $n(r)$  relation   of
\cite{Chu1995}   and   the   $\Lambda_X(T,Z)$   estimated   from   the
observations (see Table \ref{tab_diffuse}) we find

\begin{equation}\label{eq_bubble_lx}
L_X=  2.6  \times 10^{34}\,  I  \,  L_{37}^{33/35}\, n_{0}^{17/35}  \,
t_6^{19/35} \,\,\,\rm (erg \, s^{-1}),
\end{equation}

\noindent where $L_{37}$ is the mechanical luminosity of the starburst
in units of $\rm  10^{37}\,erg\,s^{-1}$, $n_{0}$ is the number density
of  the ambient  ISM in  $\rm cm^{-3}$  and $t_6$  is the  age  of the
starburst  in Myrs.   The dimensionless  integral $I$  has a  value of
about  2  \citep[see ][]{Chu1995}.   We  do  not  include an  explicit
dependence  on  metallicity   in  equation  \ref{eq_bubble_lx}  as  in
\cite{Chu1995}.  Such a dependence is accounted for in the calculation
of $\Lambda_X(T,Z)$  through the  normalisation of the  thermal plasma
model fit to the spectrum of the diffuse X-ray component.


For the  estimation of the  mechanical luminosity injected by  the the
winds of young stars and  supernovae explosions we use the STARBURST99
stellar evolution  synthesis model \citep{Leitherer1999, Vazquez2005}.
We  adopt a  continuous star-formation  model for  the  starburst with
a Salpeter IMF (lower  mass $1\,M_{\odot}$, upper mass $100\,M_{\odot}$,
slope 2.35) and star-formation  rate of $\rm 1\,M_{\odot} \, yr^{-1}$ and
metallicity $\rm  Z=0.008$, i.e.  similar to  the 0.5\,$\rm Z_{\odot}$
metallicity  of  MRK\,996  \citep{James2009}.   The  model  mechanical
luminosity is then scaled to the star-formation rate of MRK\,996, for which
we adopt  the value $\rm 1.3\,M_{\odot} \, yr^{-1}$.   This was estimated
from  the  integral   field  spectroscopy  observations  presented  by
\cite{James2009} by integrating the  narrow line component of the $\rm
H\alpha$ emission  line under the assumption that  the broad component
of this line is associated with shocks.  \cite{James2009} estimated an
age for the young stellar  population in MRK\,996 of 3-5\,Myr based on
the equivalent width  of the $\rm H\beta$ line.   An average starburst
age of 4\,Myr is  adopted in equation \ref{eq_bubble_lx}.  Under those
assumptions the  mechanical luminosity  from the STARBURST99  model is
$3.9\times10^{40}  \rm \,  erg \,  s^{-1} \,  cm^{-2}$ (Figure  112 of
Leitherer et al.  1999).  For the  density of the ambient ISM James et
al. (2009)  estimate $n_{0}=1-40\, \rm cm^{-3}$ for  the outer regions
of MRK\,996 based on  the [S\,II] $\lambda6716/ \lambda6731$ intensity
ratio.  In ionized nebulae  however, the [S\,II] emission is typically
associated with lower ionization gas  which would be expected to be of
higher density than the average. It is therefore likely that the lower
end of the $n_{0}$ range determined by James et al.  (2009) represents
the  mean ISM density  into which  the superbubble  expands.  Adopting
$n_{0}=1\,  \rm  cm^{-3}$  we  predict  an X-ray  luminosity  of  $\rm
3\times10^{38} \rm  \, erg \,  s^{-1}$, larger than the  observed one,
$\rm 1.1\times10^{38} \rm \, erg \, s^{-1}$.  However, given the large
uncertainties in the model parameters (e.g.  a slightly younger age of
3\,Myr  yields $L_X=7\times10^{37}  \rm  \, erg  \,  s^{-1}$) and  the
observables (e.g.  $\Lambda_X(T,Z)$, metallicity of the X-ray emitting
gas, ISM density)  we conclude that the observed  $L_X$ of the diffuse
component  is  broadly  consistent  with the  expanding  bubble  model
prediction.

We  caution that  the diffuse  component could  be the  result  of the
combined emission of high mass  X-ray binaries (HMXRBs) that lie below
the  sensitivity limit  of the  Chandra observations.   We  assess the
contribution  of such  sources  by integrating  the global  luminosity
function   of    HMXRBs   in   starburst    galaxies   determined   by
\cite{Grimm2003}.  A star-formation rate of $\rm 1.3\,M_{\odot}/yr$ is
adopted for MRK\,996 \citep{James2009}  and the integration is carried
out between  $\rm 10^{37}$  and $\rm 10^{38}  \, erg \,  s^{-1}$.  The
upper  limit   is  the  approximate  0.5-10\,keV   band  point  source
sensitivity limit  of the Chandra  observations.  We estimate  a total
X-ray luminosity for HMXRBs below the Chandra data detection threshold
of $\rm \approx  10^{39} \, erg \, s^{-1}$.   In principle the diffuse
X-ray component could be attributed  to HMXRBs.  The X-ray spectrum of
the diffuse  component is  not inconsistent with  this interpretation.
The best-fit spectral index of the power-law component is $2.2\pm0.7$.
Although  HMXRBs typically  have flatter  X-ray  spectra, $\approx1.2$
\citep[e.g.][]{White_Swank_Holt1983}, systems with spectral indices as
flat       as      $\Gamma\approx2$      have       been      observed
\citep[e.g.][]{Sasaki2003}.   Although  it   is  hard  to  reject  the
possibility of  X-ray binaries dominating the  diffuse component there
are arguments  against this interpretation.  The age  of the starburst
in  MRK\,996  is $3-5$\,Myr  \citep{James2009},  while  models of  the
evolution      of     the      X-ray     luminosity      of     HMXRBs
\citep[e.g.][]{Sipior2003PhD,Linden2010}  predict  that these  sources
peak  at later  times,  $\approx10-20$\,Myr after  the initial  burst,
especially  in  the case  of  low  metallicity  systems.  It  is  also
interesting that the peak of the diffuse X-ray emission falls between
the nuclear starburst and the young super-star cluster candidate in the
SW  of  the nucleus  identified  by Thuan  et  al.  (1996). One  could
reasonably  assume  that  the  peak  emission is  associated  with  an
interaction  region where  the respective  stellar wind  super bubbles
from the two sources collide and heat the ISM.

\section{Conclusions}

Chandra X-ray  observations are used to  search for an AGN  in the BCD
MRK\,996, which could explain the high ionisation line [OIV]\,25.89$\,
\rm  \mu  m$ observed  in  the mid-IR  spectrum  of  the galaxy.   The
estimated  upper limit  for the  X-ray luminosity  of a  nuclear point
source  in   MRK\,996  is  inconsistent  with   AGN  ionisation  being
responsible  for the  excitation  of the  [OIV]\,25.89$\,  \rm \mu  m$
line.  Shock excitation  must be  responsible for  this line.  We find
direct evidence for  shocks by studying the diffuse  X-ray emission of
MRK\,996.   The properties  of this  component are  broadly consistent
with shock  heating from stellar  winds and supernovae  explosions.  A
tight upper  limit of $\rm  \la 10^4 \lambda^{-1}\, M_{\odot}$  is also
set for the posited black hole mass of MRK\,996.

\section{Acknowledgments}
AG acknowledges  financial support from  the Marie-Curie Reintegration
Grant  PERG03-GA-2008-230644. YGT  acknowledges the  award of  a Marie
Curie  intra-European  Fellowship  within  the 7$^{\rm  th}$  European
Community  Framework Programme (grant  agreement PIEF-GA-2009-236486).
Support for  this work  was provided by  the National  Aeronautics and
Space Administration  through Chandra Award Number  11700503 issued by
the  Chandra  X-ray  Observatory  Center,  which is  operated  by  the
Smithsonian  Astrophysical  Observatory  for  and  on  behalf  of  the
National Aeronautics Space Administration under contract NAS8-03060.

\bibliography{mybib}{}
\bibliographystyle{mn2e}

\end{document}